\documentclass[sigconf]{cidr} 

\AtBeginDocument{%
  \providecommand\BibTeX{{%
    \normalfont B\kern-0.5em{\scshape i\kern-0.25em b}\kern-0.8em\TeX}}}

\setcopyright{acmcopyright}
\copyrightyear{2022}
\acmYear{2022}

\acmConference[CIDR 2023]{13th Annual Conference on Innovative Data Systems Research}{January 8-11, 2023}{Amsterdam, The Netherlands.}
\usepackage{color}
\usepackage{algorithmic}
\usepackage{graphicx}
\usepackage{textcomp}
\usepackage{xcolor}
\def\BibTeX{{\rm B\kern-.05em{\sc i\kern-.025em b}\kern-.08em
    T\kern-.1667em\lower.7ex\hbox{E}\kern-.125emX}}

\usepackage{subcaption}
\usepackage{mwe}

\usepackage{float}
\usepackage{hyperref}

\usepackage[english]{babel}

\usepackage{etoolbox}
\makeatletter
\patchcmd{\maketitle}{\@copyrightspace}{}{}{}
\makeatother

\iftrue
\makeatletter
\def\@copyrightspace{\relax}
\makeatother
\fi

\usepackage[numbers]{natbib}
\usepackage{ragged2e}

\begin{document}

\title{Deep Lake: a Lakehouse for Deep Learning}

\author{Sasun Hambardzumyan, Abhinav Tuli, Levon Ghukasyan, Fariz Rahman, Hrant Topchyan, David Isayan, Mark McQuade, Mikayel Harutyunyan, Tatevik Hakobyan, Ivo Stranic, Davit Buniatyan}

\iftrue \email{team@activeloop.ai} \fi
\affiliation{%
  \institution{Activeloop}
  \streetaddress{238 Castro street}
  \city{Mountain View}
  \state{CA}
  \country{USA}
  \postcode{94043}
}

\renewcommand{\shortauthors}{}

\begin{abstract}
Traditional data lakes provide critical data infrastructure for analytical workloads by enabling time travel, running SQL queries, ingesting data with ACID transactions, and visualizing petabyte-scale datasets on cloud storage. They allow organizations to break down data silos, unlock data-driven decision-making, improve operational efficiency, and reduce costs. However, as deep learning usage increases, traditional data lakes are not well-designed for applications such as natural language processing (NLP), audio processing, computer vision, and applications involving non-tabular datasets. 

This paper presents Deep Lake, an open-source lakehouse for deep learning applications developed at Activeloop\iftrue\footnote{Source code available: https://github.com/activeloopai/deeplake}\footnote{ Documentation available at https://docs.deeplake.ai}\fi. Deep Lake maintains the benefits of a vanilla data lake with one key difference: it stores complex data, such as images, videos, annotations, as well as tabular data, in the form of tensors and rapidly streams the data over the network to (a) Tensor Query Language, (b) in-browser visualization engine, or (c) deep learning frameworks without sacrificing GPU utilization. Datasets stored in Deep Lake can be accessed from PyTorch  \cite{pytorch}, TensorFlow \cite{TensorFlow}, JAX \cite{jax}, and integrate with numerous MLOps tools.
\end{abstract}

\iftrue
\fi
\maketitle

\section{Introduction}

A data lake is a central repository that allows organizations to store structured, unstructured, and semi-structured data in one place. Data lakes provide a better way to manage, govern, and analyze data. In addition, they provide a way to break data silos and gain insights previously hidden in disparate data sources. First-generation data lakes traditionally collected data into distributed storage systems such as HDFS \cite{hdfs} or AWS S3 \cite{s3}. Unorganized collections of the data turned data lakes into "data swamps", which gave rise to the second-generation data lakes led by Delta, Iceberg, and Hudi \cite{delta, iceberg, hudi}. They strictly operate on top of standardized structured formats such as Parquet, ORC, Avro  \cite{parquet, orc, avro} and provide features like time travel, ACID transactions, and schema evolution. Data lakes directly integrate with query engines such as Presto, Athena, Hive, and Photon \cite{presto, athena, hive, photon} to run analytical queries. Additionally, they connect to frameworks like Hadoop, Spark, and Airflow \cite{hadoop, zaharia2010spark, airflow} for ETL pipeline maintenance. In its turn, the integration between data lakes and query engines with clear compute and storage separation resulted in the emergence of systems like Lakehouse \cite{lakehouse} that serve as an alternative to data warehouses, including Snowflake, BigQuery, Redshift, and Clickhouse \cite{snowflake, bigquery, redshift,  clickhouse}. 

\iftrue

\begin{figure}  
    \centering
    \includegraphics[width=0.52\textwidth]{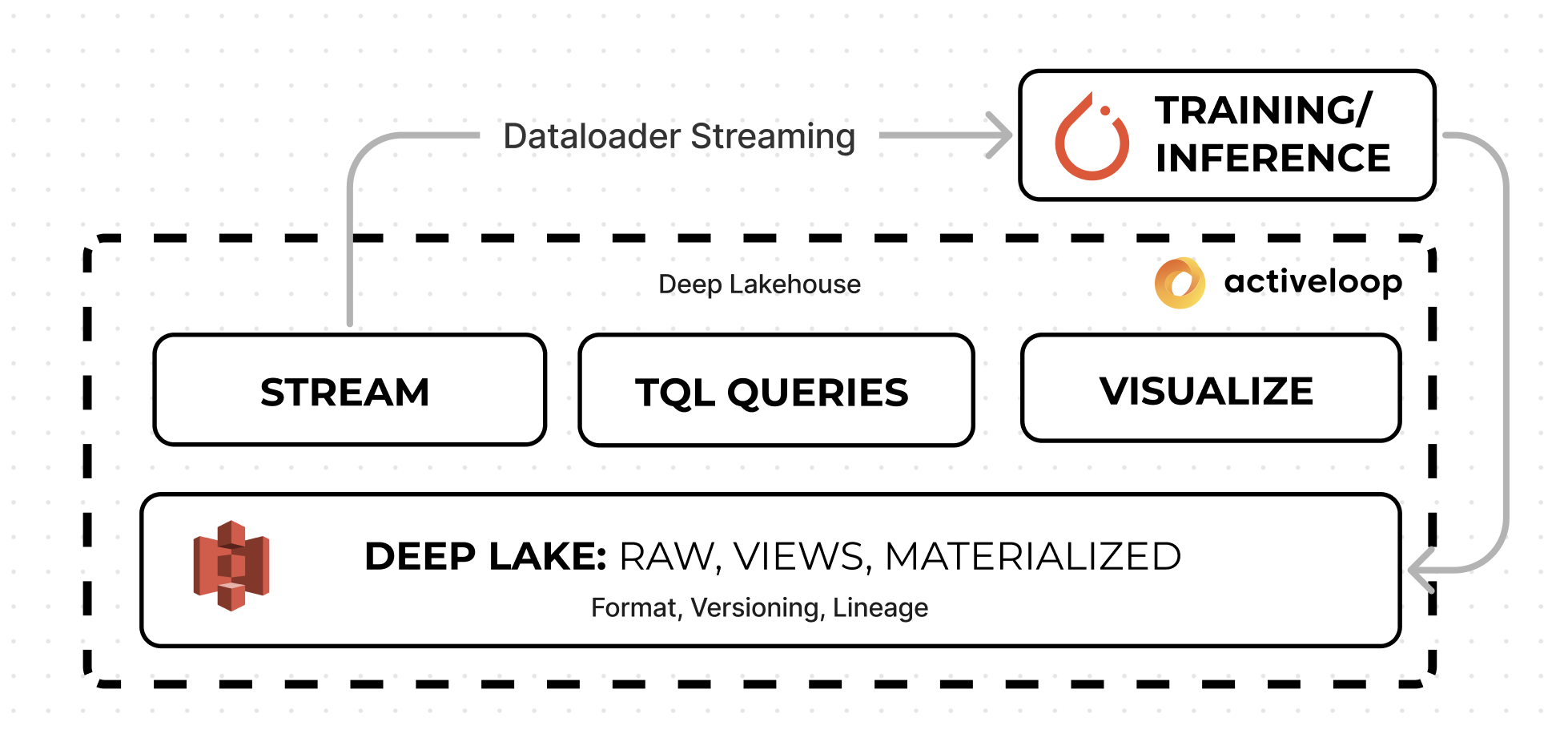}
    \caption{Deep Lake Architecture overview interfacing with deep learning frameworks.}
    \label{fig:lakehouse}
\end{figure}

Over the past decade, deep learning has outpaced traditional machine learning techniques involving unstructured and complex data such as text, images, videos, and audio \cite{krizhevsky2012imagenet, lecun2015deep, goodfellow2016deep, zhang2015character, mikolov2013efficient, bahdanau2014neural, redmon2016you, oord2016wavenet}. Not only did deep learning systems outgrow traditional techniques, but they also achieved super-human accuracy in applications such as cancer detection from X-ray images, anatomical reconstruction of human neural cells, playing games, driving cars, unfolding proteins, and generating images \cite{rajpurkar2017chexnet, lee2017superhuman, silver2018general, huang2020survey, tunyasuvunakool2021highly}. Large language models with transformer-based architectures achieved state-of-the-art results across translation, reasoning, summarization, and text completion tasks \cite{vaswani2017attention, devlin2018bert, yang2019xlnet, brown2020language}. Large multi-modal networks embed unstructured data into vectors for cross-modal search \cite{baevski2022data2vec, radford2021learning}. Moreover, they are used to generate photo-realistic images from text \cite{ramesh2021zero, saharia2022photorealistic}.

Although one of the primary contributors to the success of deep learning models has been the availability of large datasets such as CoCo (330K images), ImageNet (1.2M images), Oscar (multilingual text corpus), and LAION (400M and 5B images) \cite{lin2014microsoft, deng2009imagenet, suarez2019asynchronous, schuhmann2021laion}, it does not have a well-established data infrastructure blueprint similar to traditional analytical workloads to support such scale. On the other hand, Modern Data Stack (MDS) lacks the features required to deploy performant deep learning-based solutions so organizations opt to develop in-house systems.

\begin{figure*}
    \centering
    \includegraphics[width=\textwidth]{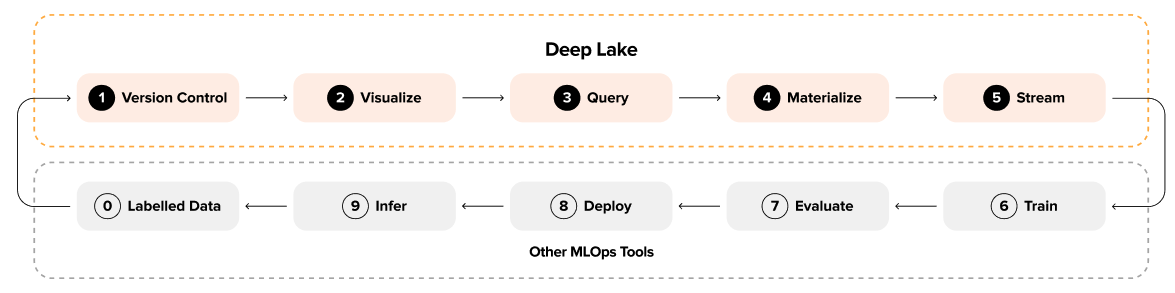}
    \caption{Machine Learning Loop with Deep Lake}
    \label{fig:worfklow}
\end{figure*}

In this paper, we introduce Deep Lake, a lakehouse specialized for deep learning workloads. Deep Lake retains the main benefits of a traditional data lake with one notable distinction: it stores complex data, such as images, videos, annotations, and tabular data, as tensors and rapidly streams the data to deep learning frameworks over the network without sacrificing GPU utilization. Furthermore, it provides native interoperability between deep learning frameworks such as PyTorch, TensorFlow, and JAX \cite{pytorch, TensorFlow, jax}.  

The main technical contributions of this paper include: 

\begin{itemize}
  \item \textit{Tensor Storage Format} that stores dynamically shaped arrays on object storage;
  \item \textit{Streaming Dataloader} that schedules fetching, decompression, and user-defined transformations, optimizing data transfer throughput to GPUs for deep learning;
  \item \textit{Tensor Query Language} running SQL-like operations on top of multi-dimensional array data;
  \item \textit{In-browser visualization engine} that streams data from object storage and renders it in the browser using WebGL.
\end{itemize}

\iftrue
The remainder of this paper unfolds as follows. We begin by considering current challenges in deep learning on unstructured data. Next, we present the Tensor Storage Format (TSF) with its key concepts. Furthermore, we discuss Deep Lake's capabilities and applications within the ML cycle. Next, we provide performance experiments and discuss the results. Finally, we review related work, list possible limitations, and conclude.
\fi

\section{Current Challenges}
\iftrue In this section, we discuss the current and historical challenges of unstructured or complex data management.\fi

\subsection{Complex Data Types in a Databases}

It is generally not recommended to store binary data, such as images, directly in a database. This is because databases are not optimized for storing and serving large files and can cause performance issues. In addition, binary data does not fit well into a database's structured format, making it difficult to query and manipulate. This can lead to slow load times for users. Databases are typically more expensive to operate and maintain than other types of storage, such as file systems or cloud storage services. Therefore, storing large amounts of binary data in a database can be more costly than other storage solutions.

\subsection{Complex Data Along with Tabular Formats}

Increases in large-scale analytical and BI workloads motivated the development of compressed structured formats like Parquet, ORC, Avro, or transient in-memory formats like Arrow \cite{parquet, orc, avro, arrow}. As tabular formats gained adoption, attempts to extend those formats, such as Petastorm \cite{Petastorm} or Feather \cite{feather} for deep learning, have emerged. \label{marker} To the best of our knowledge, these formats have yet to gain wide adoption. This approach primarily benefits from native integrations with Modern Data Stack (MDS). However, as discussed previously, upstream tools require fundamental modifications to adapt to deep learning applications.

\subsection{Object Storage for Deep Learning}

The current cloud-native choice for storing large unstructured datasets is object storage such as AWS S3 \cite{s3}, Google Cloud Storage (GCS) \cite{gcs}, or MinIO \cite{minio}. Object storage does offer three main benefits over distributed network file systems. They are (a) cost-efficient, (b) scalable, and (c) serve as a format-agnostic repository. However, cloud storages are not without drawbacks. Firstly, they introduce significant latency overhead, especially when iterating over many small files such as text or JSON. Next, unstructured data ingestion without metadata control can produce "data swamps". Furthermore, object storage has built-in version control; it is rarely used in data science workflows. Lastly, data on object storage gets copied to a virtual machine before training, thus resulting in storage overhead and additional costs.

\subsection{Second Generation of Data Lakes}

The second-generation data lakes led by Delta, Iceberg, Hudi \cite{delta, iceberg, hudi} extend object storage by managing tabular format files with the following primary properties. 

\begin{enumerate}
  \item \textit{Update operations}: inserting or deleting a row on top of a tabular format file.
  \item \textit{Streaming}: downstream data ingestion with ACID properties and upstream integration with query engine exposing SQL interface. 
  \item \textit{Schema evolution}: evolving columnar structure while preserving backward compatibility.
  \item \textit{Time travel and audit log trailing}: preserving historical state with rollback property where queries can be reproducible. Also, support for row-level control on data lineage. 
  \item \textit{Layout optimization}: Built-in feature to optimize file sizes and data compaction with custom ordering support. Significantly speeds up querying.
\end{enumerate}

However, second-generation data lakes are still bound by the limitations of the inherent data formats to be used in deep learning, as previously discussed in section \ref{marker}. Hence in this paper, we extend the second generation of data lake capabilities for deep learning use cases by rethinking the format and upstream features, including querying, visualization, and native integration to deep learning frameworks to complete the ML lifecycle as shown in Fig. \ref{fig:worfklow}.

\section{Tensor Storage Format}

Deep Lake datasets follow columnar storage architecture, with tensors as columns, as shown in Fig. \ref{fig:layout}. Each tensor is a collection of \emph{chunks} - binary blobs that contain the data samples. An index map associated with each tensor helps find the right chunk and index of the sample within that chunk for a given sample index.

\subsection{Dataset}

A sample in a dataset represents a single row indexed across parallel tensors. As opposed to a document storage format, sample elements are logically independent, which enables partial access to samples for running performant queries or streaming selected tensors over the network to the GPU training instances. Multiple tensors can be grouped. Groups implement syntactic nesting and define how tensors are related to each other. Syntactic nesting avoids the format complication for hierarchical memory layout. Changes to the dataset's schema are also tracked over time with version control, similar to dataset content changes.

\subsection{Tensors}

Tensors are typed and can be appended or modified in-place. Default access to an index or a set of indices returns the data as NumPy arrays \cite{oliphant2006guide}. Instead of storing 1-D data as seen in Parquet \cite{parquet} or series in Arrow \cite{arrow}, tensors can accommodate n-dimensional data, where typically the first dimension corresponds to the index or batch dimension. Tensors can contain dynamically shaped arrays, also called ragged tensors, as opposed to other statically chunked array formats such as Zarr \cite{zarr}.

\subsection{Types}

Htype defines the expectations on samples in a tensor such as data type (\textit{dtype} as seen in NumPy \cite{oliphant2006guide}), shape, number of dimensions, or compression.  Typed tensors make interacting with deep learning frameworks straightforward and enable sanity checks and efficient memory layout. By inheriting from a generic tensor htype, we can construct types such as \textit{image}, \textit{video}, \textit{audio}, \textit{bbox}, \textit{dicom}, and others. For example, a tensor with \textit{image} htype would expect samples being appended to it to have \textit{dtype} as \textit{uint8} and shape length 3 (i.e. width, height and number of channels). We further expand on the notion of htypes allowing for meta types that support storing image sequences in tensors (\textit{sequence[image]}), referencing to remotely stored images, while maintaining the regular behavior of a \textit{image} tensor (\textit{link[image]}), or even possible cross-format support.

\subsection{Memory Layout}

\begin{figure}
    \centering
    \includegraphics[width=0.52\textwidth]{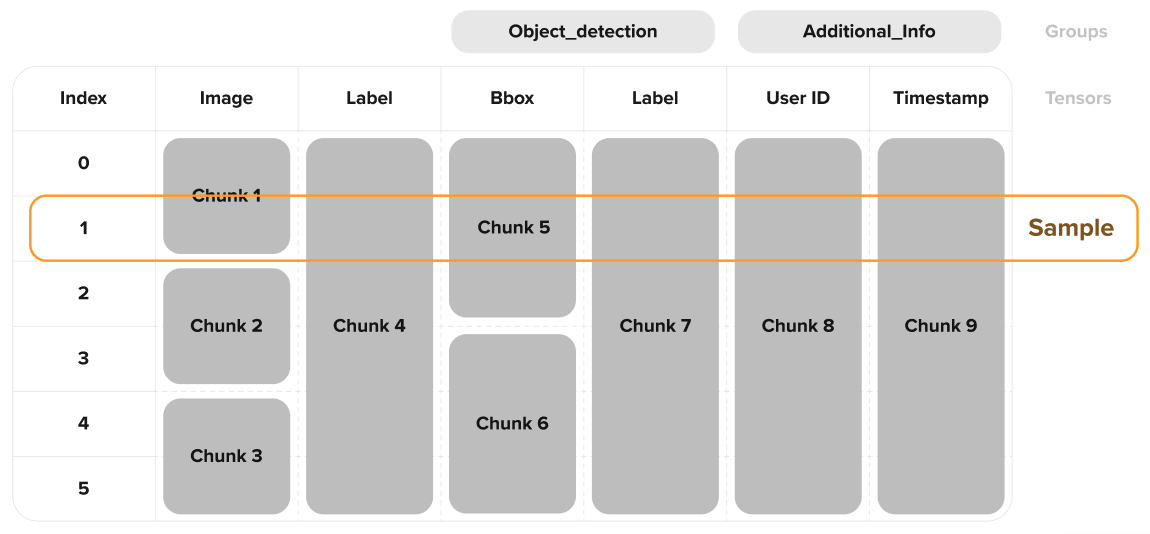}
    \caption{How each sample (row) is stored in a set of columnar tensors with dynamically sized chunks}
    \label{fig:layout}
\end{figure}
A Deep Lake dataset contains a provenance file in JSON format and folders per tensor. A tensor contains chunks, chunk encoder, tile encoder, and tensor metadata. Tensors can be optionally hidden. For instance, hidden tensors can be used to maintain down-sampled versions of images or preserve shape information for fast queries.

Tensors are stored in chunks at the storage level. While statically (inferred) shaped chunking avoids maintaining a chunk map table, it introduces significant user overhead during the specification of the tensor, custom compression usage limitations, underutilized storage for dynamically shaped tensors, and post-processing inefficiencies. Deep Lake chunks are constructed based on the lower and upper bound of the chunk size to fit a limited number of samples. This comes with a trade-off of having a compressed index map that preserves the sample index to chunk id mapping per tensor while enabling chunk sizes in the range optimal for streaming while accommodating mixed-shape samples. One could consider the approach taken in this paper as an optimized trade-off between file system page map and compute-defined map-less array storage system. For practical reasons, a single chunk encoder can be scaled to billions of images while maintaining a 150MB chunk encoder per 1PB tensor data. Further scaling can be introduced by sharding the chunk encoder. Chunks contain header information such as byte ranges, shapes of the samples, and the sample data itself. If a sample is larger than the upper bound chunk size, which is the case for large aerial or microscopy images, the sample is tiled into chunks across spatial dimensions. The only exception to tiling is videos. Videos are preserved due to efficient frame mapping to indices,  key-frame-only decompression, and range-based requests while streaming. 

\subsection{Access Patterns}

The tensor storage format is optimized for deep learning training and inference, including sequential and random access. Sequential access is used for running scan queries, transforming tensors into other tensors, or running inference. Random access use cases include multiple annotators writing labels to the same image or models storing back predictions along with the dataset. While the strict mode is disabled, out-of-the-bounds indices of a tensor can be assigned, thus accommodating sparse tensors. However, random assignment over time will produce inefficiently stored data chunks. To fix the data layout, we implement an on-the-fly re-chunking algorithm to optimize the data layout. One of the key access patterns of Deep Lake is shuffled stream access for training machine learning models. It requires random or custom order access while streaming chunks into the training process. This is achieved by involving range-based requests to access sub-elements inside chunks, running complex queries before training to determine the order, and maintaining a buffer cache of fetched and unutilized data. This avoids having a separate compute cluster for running shuffling algorithm  \cite{luan2022exoshuffle}.

Each tensor has its own chunks, and the default chunk size is 8MB. A single chunk consists of data from multiple indices when the individual data points (image, label, annotation, etc.) are smaller than the chunk size. Conversely, when individual data points are larger than the chunk size, the data is split among multiple chunks (tiling). Exceptions to chunking logic are video data.

Deep Lake format is optimized for maximizing throughput to GPU processing. It includes CPU pre-fetching, decompression or decoding, transformations, and GPU memory transfer in a deep learning framework's expected layout.


\subsection{Storage Providers}

Deep Lake can be plugged into any storage provider, including object storages such as AWS S3 \cite{s3}, Google Cloud Storage (GCS) \cite{gcs},  POSIX compatible file systems, or local in-memory storage. Moreover, it constructs memory caching by chaining various storage providers together, for instance - the Least Recently Used (LRU) cache of remote S3 storage with local in-memory data. 

\section{Deep Lake System Overview}
As shown in Fig. \ref{fig:lakehouse}, Deep Lake stores raw data and views in object storage such as S3 and materializes datasets with full lineage. Streaming, Tensor Query Language queries, and Visualization engine execute along with either deep learning compute or on the browser without requiring external managed or centralized service.

\subsection{Ingestion} 

\subsubsection{Extract}
Sometimes metadata might already reside in a relational database. We additionally built an ETL destination connector using Airbyte\footnote{Source code available: https://github.com/activeloopai/airbyte on the branch @feature/connector/deeplake} \cite{airbyte}. The framework allows plugging into any supported data source, including SQL/NoSQL databases, data lakes, or data warehouses, and synchronizing the data into Deep Lake. Connector protocol transforms the data into a columnar format.

\subsubsection{Transform}
To significantly accelerate data processing workflows and free users from worrying about the chunk layout, 
Deep Lake provides an option to execute python transformations in parallel. The transformation takes in a dataset, sample-wise iterates across the first dimension,  and outputs a new dataset. A user defined python function expects two required arguments \(sample\_in\), \(sample\_out\) and is decorated with \(@deeplake.compute\). A single \(sample\_in\) can dynamically create multiple \(sample\_outs\). It enables both one-to-one and one-to-many transformations. The transformation can also be applied in place without creating a new dataset. Behind the scenes, the scheduler batches sample-wise transformations operating on nearby chunks and schedule them on a process pool. Optionally, the compute can be delegated to a Ray cluster \cite{moritz2018ray}. Instead of defining an input dataset, the user can provide an arbitrary iterator with custom objects to create ingestion workflows. Users can also stack together multiple transformations and define complex pipelines.

\subsection{Version Control}

\iftrue
\begin{figure}
    \centering
    \includegraphics[width=0.5\textwidth]{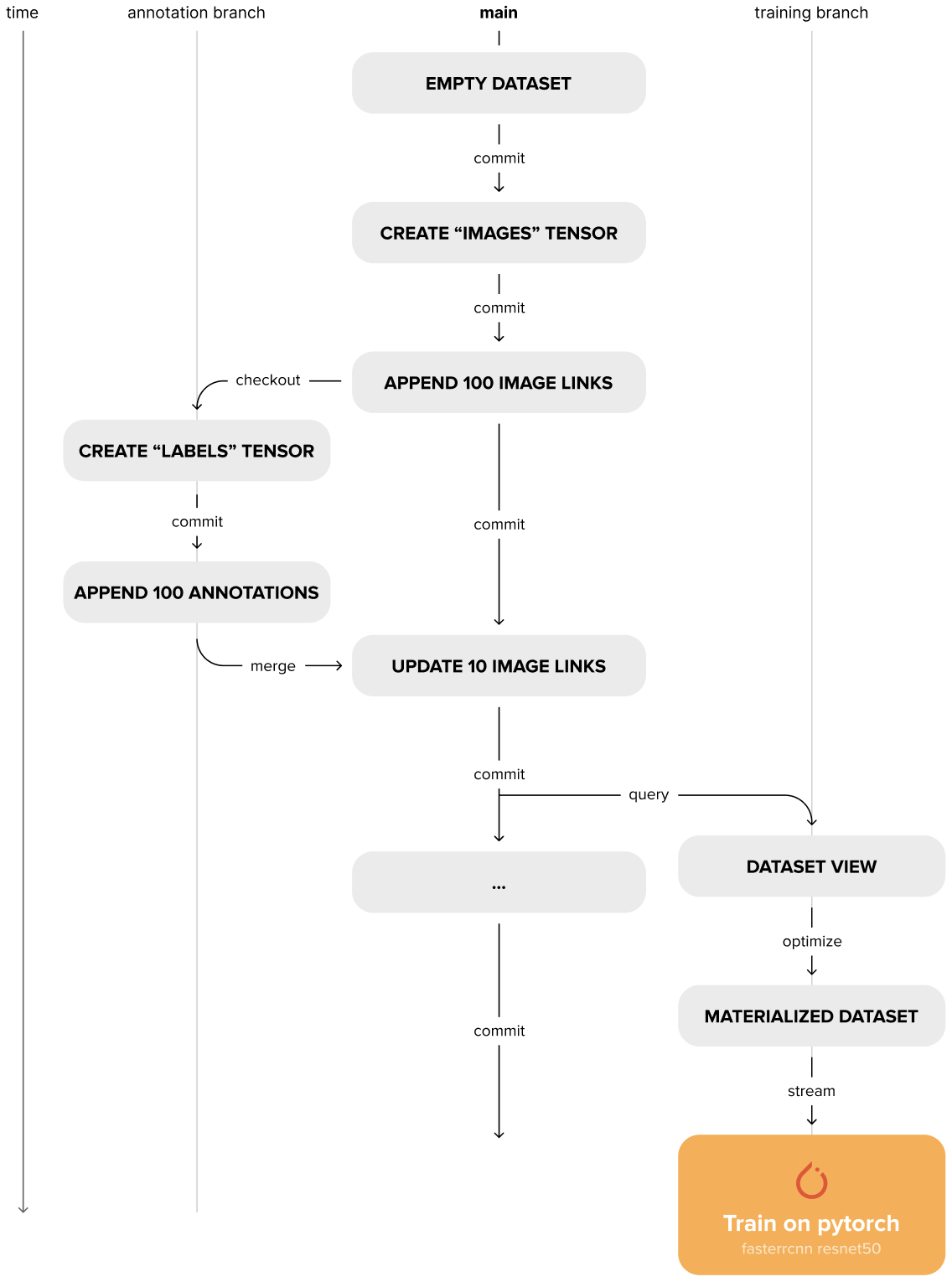}
    \caption{Version History of Evolving Deep Lake Dataset from empty till materialized view}
    \label{fig:version_workflow}
\end{figure}
\fi

Deep Lake also addresses the need for the reproducibility of experiments and compliance with a complete data lineage. Different versions of the dataset exist in the same storage, separated by sub-directories. Each sub-directory acts as an independent dataset with its metadata files. Unlike a non-versioned dataset, these sub-directories only contain chunks modified in the particular version, along with a corresponding chunk\_set per tensor containing the names of all the modified chunks. A version control info file present at the root of the directory keeps track of the relationship between these versions as a branching version-control tree. While accessing any chunk of a tensor at a particular version, the version control tree is traversed starting from the current commit, heading towards the first commit. During the traversal, the chunk set of each version is checked for the existence of the required chunk. If the chunk is found, the traversal is stopped, and data is retrieved. For keeping track of differences across versions, for each version, a commit diff file is also stored per tensor. This makes it faster to compare across versions and branches. Moreover, the ids of samples are generated and stored during the dataset population. This is important for keeping track of the same samples during merge operations. Deep Lake's version control interface is the Python API, which enables machine learning engineers to version their datasets within their data processing scripts without switching back and forth from the CLI. It supports the following commands:
\begin{itemize}
\item \textit{Commit}: creates an immutable snapshot of the current state of the dataset.
\item \textit{Checkout}: checks out to an existing branch/commit or creates a new branch if one doesn't exist.
\item \textit{Diff}: compares the differences between 2 versions of the dataset.
\item \textit{Merge}: merges two different versions of the dataset, resolving conflicts according to the policy defined by the user. 
\end{itemize}

\subsection{Visualization of Tensors}

Data visualization is a crucial part of ML workflows, especially when the data is hard to parse analytically. Fast and seamless visualization allows faster data collection, annotation, quality inspection, and training iterations. The Deep Lake visualizer engine provides a web interface for visualizing large-scale data directly from the source. It considers \textit{htype} of the tensors to determine the best layout for visualization. Primary tensors, such as \textit{image}, \textit{video} and \textit{audio} are displayed first, while secondary data and annotations, such as \textit{text}, \textit{class\_label}, \textit{bbox} and \textit{binary\_mask} are overlayed. The visualizer also considers the meta type information, such as \textit{sequence} to provide a sequential view of the data, where sequences can be played and jump to the specific position of the sequence without fetching the whole data, which is relevant for video or audio use cases. Visualizer addresses critical needs in ML workflows, enabling users to understand and troubleshoot the data, depict its evolution, compare predictions to ground truth or display multiple sequences of images (e.g., camera images and disparity maps) side-by-side.

\subsection{Tensor Query Language}

Querying and balancing datasets is a common step in training deep learning workflows. Typically, this is achieved inside a dataloader using sampling strategies or separate pre-processing steps to sub-select the dataset. On the other hand, traditional data lakes connect to external analytical query engines \cite{photon} and stream Dataframes to data science workflows. To resolve the gap between the format and fast access to the specific data, we provide an embedded SQL-like query engine implemented in C++ called Tensor Query Language (TQL). An example query is shown at Fig. \ref{fig:query_example}. While SQL parser has been extended from Hyrise \cite{hyrise} to design Tensor Query Language, we implemented our planner and execution engine that can optionally delegate computation to external tensor computation frameworks. The query plan generates a computational graph of tensor operations. Then the scheduler, executes the query graph. Execution of the query can be delegated to external tensor computation frameworks such as PyTorch \cite{pytorch} or XLA \cite{XLA} and efficiently utilize underlying accelerated hardware. In addition to standard SQL features, TQL also implements numeric computation. There are two main reasons for implementing a new query language. First, traditional SQL does not support multidimensional array operations such as computing the mean of the image pixels or projecting arrays on a specific dimension. TQL solves this by adding Python/NumPy-style indexing, slicing of arrays, and providing a large set of convenience functions to work with arrays, many of which are common operations supported in NumPy. Second, TQL enables deeper integration of the query with other features of the Deep Lake, such as version control, streaming engine, and visualization. For example, TQL allows querying data on specific versions or potentially across multiple versions of a dataset. TQL also supports specific instructions to customize the visualization of the query result or seamless integration with the dataloader for filtered streaming. The embedded query engine runs along with the client either while training a model on a remote compute instance or in-browser compiled over WebAssembly. TQL extends SQL with numeric computations on top of multi-dimensional columns. It constructs views of datasets, which can be visualized or directly streamed to deep learning frameworks. Query views, however, can be sparse, which can affect streaming performance.

\begin{figure}
    \centering    
\begin{verbatim}
SELECT 
  images[100:500, 100:500, 0:2] as crop, 
  NORMALIZE(
    boxes, 
    [100, 100, 400, 400]) as box
FROM 
  dataset 
WHERE IOU(boxes, "training/boxes") > 0.95
ORDER BY IOU(boxes, "training/boxes")
ARRANGE BY labels
\end{verbatim}
\caption{An example query that arranges cropped images ordered by bounding boxes predictions error measured over user-defined function IOU (Intersection over Union).}
\label{fig:query_example}
\end{figure}

\subsection{Materialization}
Most of the raw data used for deep learning is stored as raw files (compressed in formats like JPEG), either locally or on the cloud. A common way to construct datasets is to preserve pointers to these raw files in a database, query this to get the required subset of data, fetch the filtered files to a machine, and then train a model iterating over files. In addition, data lineage needs to be manually maintained with a provenance file. Tensor Storage Format simplifies these steps using linked tensors - storing pointers (links/urls to one or multiple cloud providers) to the original data. The pointers within a single tensor can be connected to multiple storage providers, thus allowing users to get a consolidated view of their data present in multiple sources. All of Deep Lake’s features including queries, version control, and streaming to deep learning frameworks can be used with linked tensors. However, the performance of data streaming will not be as optimal as default tensors. A similar problem exists with sparse views created due to queries, which would be inefficiently streamed due to the chunk layout. Furthermore, materialization transforms the dataset view into an optimal layout to stream into deep learning frameworks to iterate faster. Materialization involves fetching the actual data from links or views and efficiently laying it out into chunks. Performing this step towards the end of machine learning workflows leads to minimum data duplication while ensuring optimal streaming performance and minimal data duplication, with full data lineage.

\subsection{Streaming Dataloader} 

As datasets become larger, storing and transferring over the network from a remotely distributed storage becomes inevitable. Data streaming enables training models without waiting for all of the data to be copied to a local machine. The streaming dataloader ensures data fetching, decompression, applying transformations, collation, and data handover to the training model. Deep learning dataloaders typically delegate fetching and transformation to parallel running processes to avoid synchronous computation. Then the data is transferred to the main worker through inter-process communication (IPC) which introduces memory copy overhead or uses shared memory with some reliability issues. In contrast, Deep Lake dataloader delegates highly parallel fetching and in-place decompressing in C++ per process to avoid global interpreter lock. Then, it passes the in-memory pointer to the user-defined transformation function and collates before exposing them to the training loop in deep learning native memory layout. Transformation concurrently executes in parallel when it uses only native library routine calls and releases python global interpreter lock (GIL) accordingly. As a result, we get:
\begin{itemize}
\item \textit{Performance}: Delivering data to the deep learning model fast enough so that either the GPU is fully utilized or bottlenecked by the compute.
\item \textit{Smart Scheduler}: Dynamically differentiating between CPU-intensive jobs prioritization over less-intensive.
\item \textit{Efficient Resource Allocation}: Predicting memory consumption to avoid breaking the training process due to memory overfilling.
\end{itemize}


\iftrue

\section{Machine Learning Use Cases}
In this section, we review the applications of Deep Lake.

A typical scenario in a Deep Learning application starts with 
\begin{enumerate}
    \item A raw set of files that is collected on an object storage bucket. It might include images, videos, and other types of multimedia data in their native formats such as JPEG, PNG or MP4.
    \item Any associated metadata and labels stored on a relational database. Optionally, they could be stored on the same bucket along with the raw data in a normalized tabular form such as CSV, JSON, or Parquet format.
\end{enumerate}
As shown in Fig. \ref{fig:version_workflow}, an empty Deep Lake dataset is created. Then, empty tensors are defined for storing both raw data as well as metadata. The number of tensors could be arbitrary. A basic example of an image classification task would have two tensors, 
\begin{itemize}
    \item \textit{images} tensor with htype of \(image\) and sample compression of JPEG
    \item \textit{labels} tensor with htype of \(class\_label\) and chunk compression of LZ4. 
\end{itemize}
After declaring tensors, the data can be appended to the dataset. If a raw image compression matches the tensor sample compression, the binary is directly copied into a chunk without additional decoding. Label data is extracted from a SQL query or CSV table into a categorical integer and appended into \textit{labels} tensor. \textit{labels} tensor chunks are stored using LZ4 compression. All Deep Lake data is stored in the bucket and is self-contained. After storage, the data can be accessed in a NumPy interface or as a streamable deep learning dataloader. Then, the model running on a compute machine iterates over the stream of image tensors, and stores the output of the model in a new tensor called \textit{predictions}. Furthermore, we discuss below how one can train, version control, query, and inspect the quality of a Deep Lake dataset.

\subsection{Deep Learning Model Training}

Deep learning models are trained at multiple levels in an organization, ranging from exploratory training occurring on personal computers to large-scale training that occurs on distributed machines involving many GPUs. The time and effort required to bring the data from long-term storage to the training client are often comparable to the training itself. Deep Lake solves this problem by enabling rapid streaming of data without bottlenecking the downstream training process, thus avoiding the cost and time required to duplicate data on local storage.

\subsection{Data Lineage and Version Control}

Deep learning data constantly evolve as new data is added and existing data is quality controlled. Analytical and training workloads occur in parallel while the data is changing. Hence, knowing which data version was used by a given workload is critical to understand the relationship between the data and model performance. Deep Lake enables deep learning practitioners to understand which version of their data was used in any analytical workload and to time travel across these versions if an audit is required. Since all data is mutable, it can be edited to meet compliance-related privacy requirements. Like Git for code, Deep Lake also introduces the concept of data branches, allowing experimentation and editing of data without affecting colleagues’ work.

\subsection{Data Querying and Analytics}

Training of deep learning models rarely occurs on all data collected by an organization for a particular application. Training datasets are often constructed by filtering the raw data based on conditions increasing model performance, which often includes data balancing, eliminating redundant data, or selecting data that contains specific features. Deep Lake provides the tools to query and analyze data so that deep learning engineers can create datasets yielding the highest accuracy models.

\subsection{Data Inspection and Quality Control}

Though unsupervised learning is becoming more applicable in real-world use cases, most deep learning applications still rely on supervised learning. Any supervised learning system is only as good as the quality of its data, often achieved by manual and exhaustive inspection of the data. Since this process is time-consuming, it is critical to provide the humans in the loop with tools to examine vast amounts of data very quickly. Deep Lake allows inspecting deep learning datasets of any size from the browser without any setup time or need to download data. Furthermore, the tools can be extended for comparing model results with ground truth. Combined with querying and version control, this can be applied to the iterative improvement of data to achieve the best possible model.

\fi

\section{Performance Benchmarks}

In this section, we experimentally demonstrate Deep Lake’s performance at scale from the point of ingestion into the format up to training at scale against other dataloaders and formats. We compare streaming datasets from different storage backends, and showcase performance gains and scalability while training on the cloud.

\subsection{Ingestion speed to various formats}

\begin{figure}[h]
    \centering
    \includegraphics[width=0.475\textwidth]{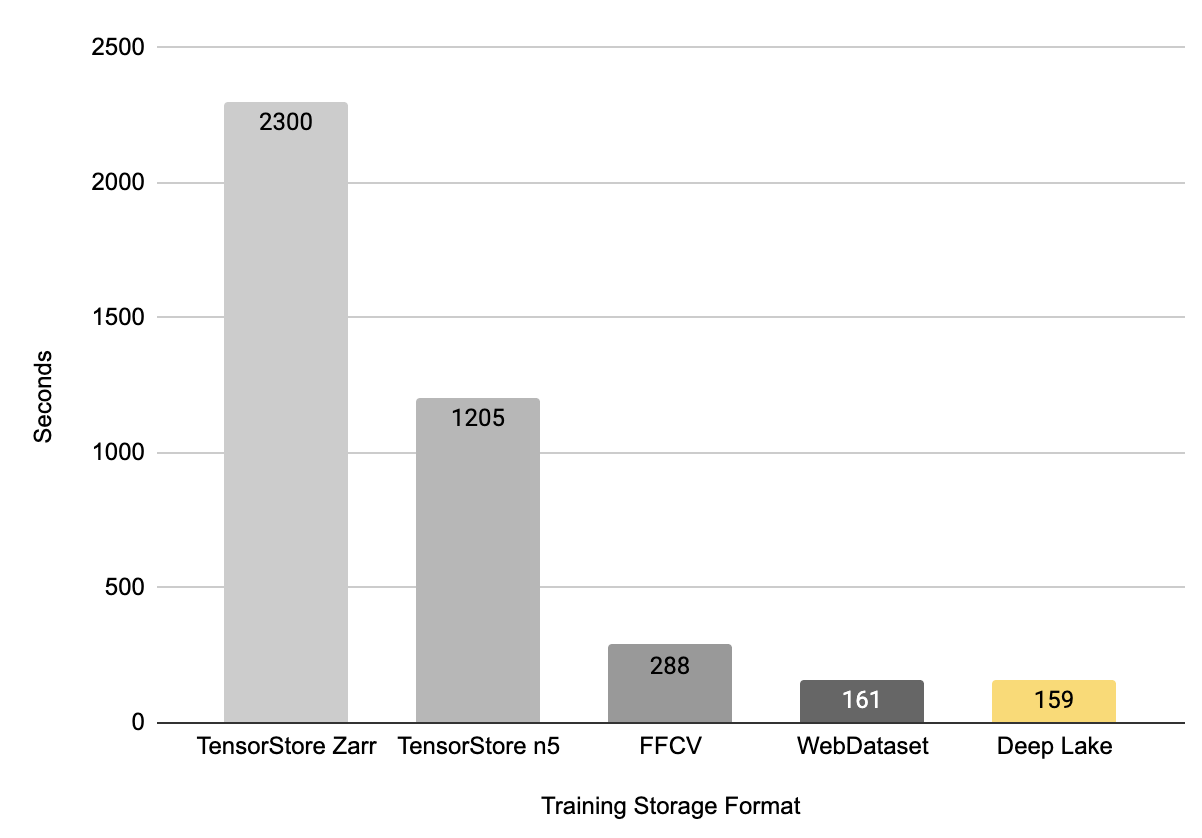}
    \caption[Network2]%
    {Ingesting 10,000 images from FFHQ \cite{ffhq} dataset into different format (lower better)}    
    \label{fig:ingestingffhq}
\end{figure}

10,000 images from FFHQ \cite{ffhq} dataset were uncompressed and stored in NumPy format. Each 1024x1024x3 raw image is a 3MB array. Then, as shown in Fig. \ref{fig:ingestingffhq} images were serially written into each format. To increase the performance, we used TensorStore \cite{tensorstore} to write to Zarr \cite{zarr} and N5 \cite{N5} formats. The experiments were done on the AWS c5.9xlarge machine. Deep Lake achieves significantly faster write performance compared to array formats and on par with binary formats such as WebDataset \cite{webdataset} and FFCV Beton \cite{ffcv}

\subsection{Comparison of local dataloaders}
\begin{figure}[h]
    \centering 
    \includegraphics[width=0.475\textwidth]{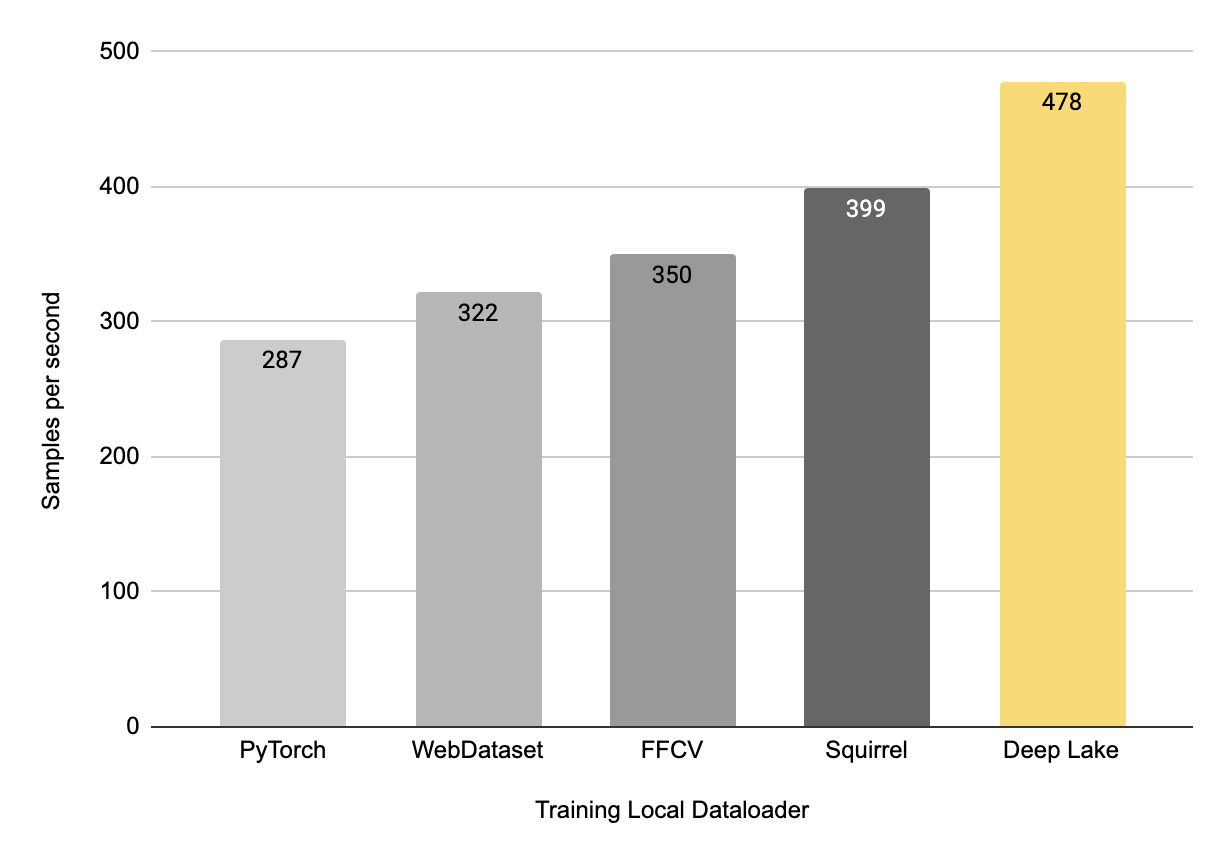}
    \caption[Network2]{Iteration speed of images against other dataloaders (higher better)}%
    \label{fig:dataloader_benchmarks}
\end{figure}
\label{marker:localdataloader} 

As shown in Fig. \ref{fig:dataloader_benchmarks} Deep Lake achieves faster data loading in a PyTorch training loop without a model. The experiment was carried out on AWS P3.2xlarge instance with one Nvidia V100 GPU card. The dataset has randomly generated 50,000 250x250x3 images stored as JPEG files. The list of libraries in which the benchmarks were carried out was Deep Lake, FFCV \cite{ffcv}, Squirrel \cite{2022squirrelcore}, Webdataset \cite{webdataset} and native PyTorch dataloader \cite{pytorch}. 

\subsection{Streamable dataloader from different locations}
\begin{figure}[h]
    \centering 
    \includegraphics[width=0.475\textwidth]{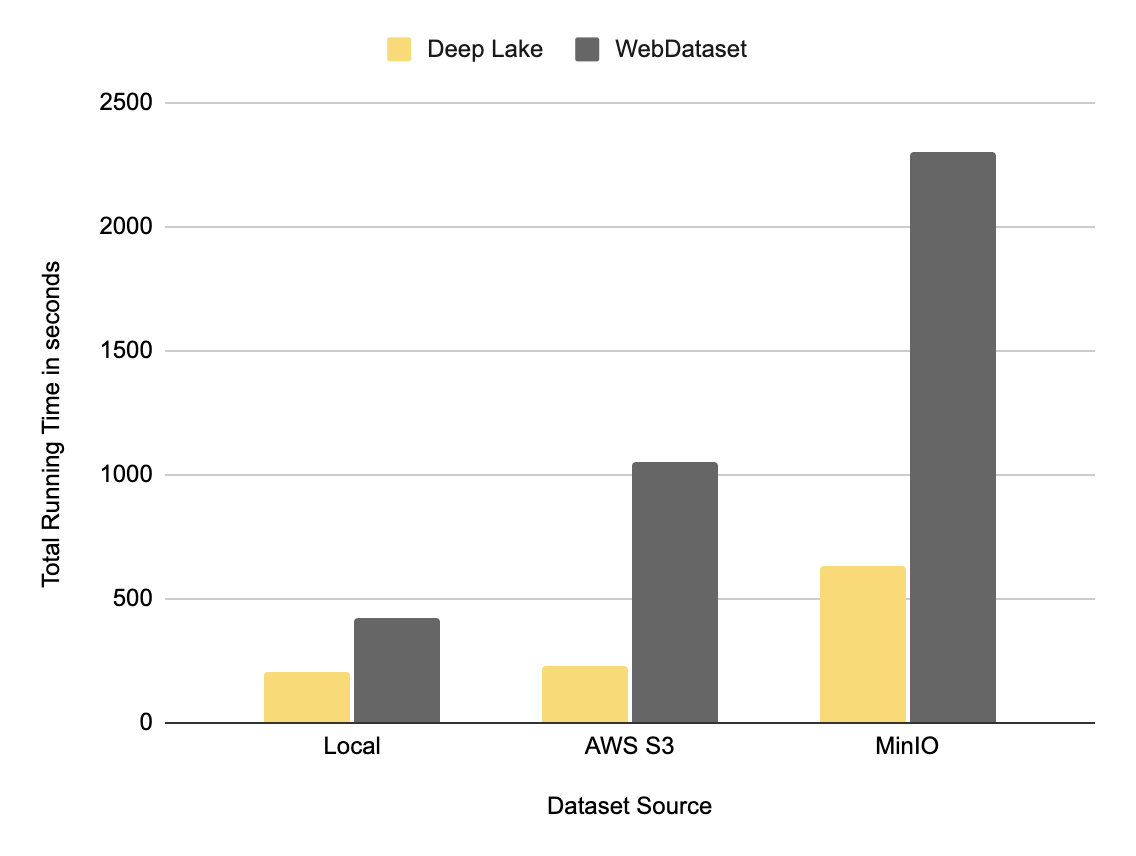}
    \caption[]%
    {Streaming from different data storage locations: Local FileSystem, AWS S3, MinIO (lower better)}    
    \label{fig:storagestreaming}
\end{figure}
In this experiment as shown in Fig. \ref{fig:storagestreaming}, we explore different storage backends for remote streaming using the same dataset as in Section \ref{marker:localdataloader}. MinIO \cite{minio} is running on another machine in a local network. Notably, Deep Lake achieves similar performance as if the data is local to the machine compared to AWS S3. Both WebDataset and Deep Lake are significantly slower while streaming the data from MinIO compared to AWS S3. For more detailed dataloader benchmarks, we would recommend an exhaustive dataloader overview study by Ofeidis et al.  \cite{ofeidis2022overview}.


\subsection{ImageNet training on the cloud}
\begin{figure}
    \centering
    \includegraphics[width=0.5\textwidth]{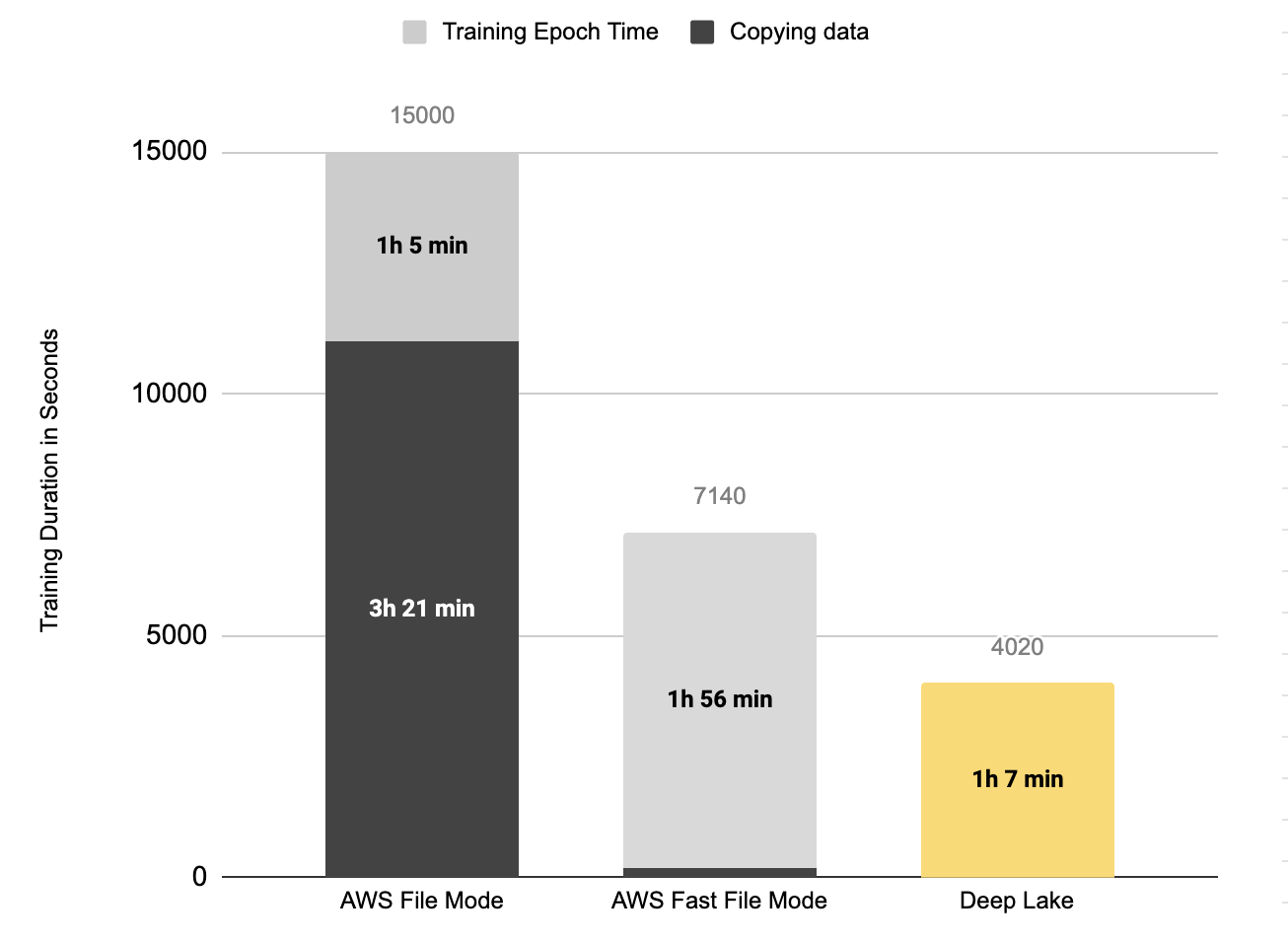}
    \caption{Training on ImageNet on an S3: AWS File Mode copies file by file from S3; Fast File Mode starts immediately with slower training; Deep Lake performs as if data is local, although it is streamed (lower better)
    }
    \label{fig:imagenet_training}
\end{figure}

Since Deep Lake was built to be cloud-first, in this and next section we demonstrate the benefits it provides for training models on the cloud. We take ImageNet dataset \cite{imagenet_cvpr09} and store it on AWS S3 \iftrue \cite{s3} \fi as original and Tensor Storage Format. The dataset contains 1.2 million images and labels in total 150GB. Deep Lake achieves virtually similar training performance as if the data were local to the machine. This saves up to 4x GPU compute time and cost as shown in Fig. \ref{fig:imagenet_training}

\subsection{Distributed training of a large multi-modal dataset}

\begin{figure}
    \centering
    \includegraphics[width=0.5\textwidth]{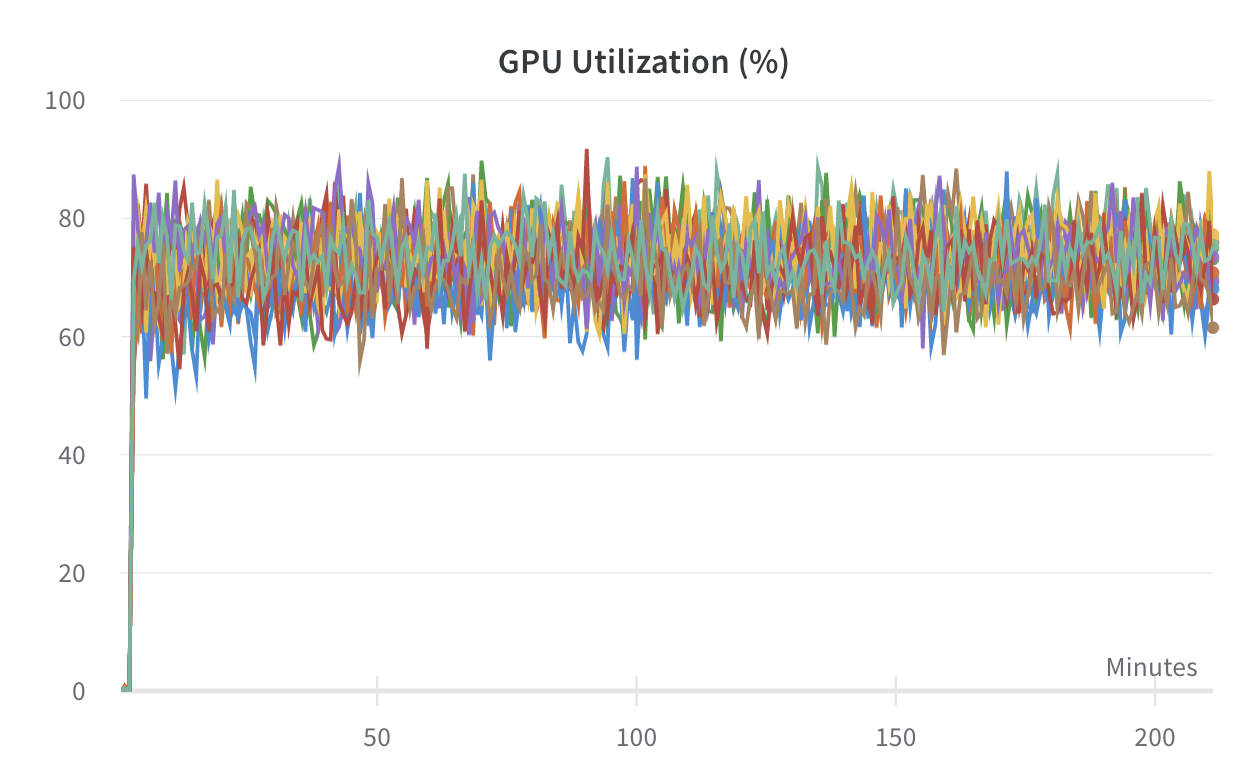}
    \caption{GPU utilization of single 16xA100 GPU machine while training 1B parameter CLIP model \cite{radford2021learning}. The dataset is LAION-400M \cite{schuhmann2021laion} streaming from AWS us-east to GCP us-central datacenter. Each color demonstrates single A100 GPU utilization over training.}
    \label{fig:laion_training}
\end{figure}

As a second experiment, we take LAION dataset \cite{schuhmannlaion} containing 400M image-text pairs and train CLIP \cite{radford2021learning}, image-text embedding model with 1 billion parameters. The original dataset is a table of Parquet files with a column of image URLs. The dataset download from the source took 100 hours, while ingestion to Tensor Storage format took only 6 hours, totaling 1.9TB in size. The dataset has been stored on AWS in the US-east region while training GPU machine in the US-central region. As shown on Fig. \ref{fig:laion_training} Deep Lake achieves high GPU utilization by streaming 5,100 images/s into 16 Nvidia A100 GPUs while without model up to 80,000 images/s per machine on the same region.

\section{Discussion and Limitations}

Deep Lake's primary use cases include (a) Deep Learning Model Training, (b) Data Lineage and Version Control, (c) Data Querying, and Analytics, (d) Data Inspection and Quality Control. We took NumPy \cite{oliphant2006guide} arrays as a fundamental block and implemented version control, streaming dataloaders, visualization engine from scratch.

\subsection{Format Design Space}

The Tensor Storage Format (TSF) is a binary file format designed specifically for storing tensors, which are multi-dimensional arrays of numerical values used in many machine learning and deep learning algorithms. The TSF format is designed to be efficient and compact, allowing for fast and efficient storage and access of tensor data. One key advantage of the TSF format is that it supports a wide range of tensor data types, including dynamically shaped tensors.

In comparison, the Parquet \cite{parquet} and Arrow \cite{arrow} formats are columnar file formats that are designed for storing and processing large analytical datasets. Unlike TSF, which is specifically designed for tensor data, Parquet and Arrow are optimized for efficient storage and querying of analytical workloads on tabular and time-series data. They use columnar storage and compression techniques to minimize storage space and improve performance, making them suitable for big data applications. However, TSF has some advantages over Parquet and Arrow when it comes to tensor data. TSF can support tensor operations and efficient streaming to deep learning frameworks.

Other tensor formats \cite{Petastorm, zarr, tensorstore, papadopoulos2016tiledb} are efficient for massively massively parallelizable workloads as they don't require coordination across chunks. Tensor Storage Format key trade-off is enabling to store dynamically shape arrays inside a tensor without padding memory footprint. For example, in computer vision it is very common to store multiple images with different shapes or videos have dynamic length. To support the flexibility, minor overhead is introduced in the form of previously discussed chunk encoder that in practice we haven't observed impact on production workloads.

\subsection{Dataloader}
Deep Lake achieves state-of-the-art results in local and remote settings, as seen in benchmarks for iterating on large images Fig. \ref{fig:dataloader_benchmarks}. Primarily, it has been faster than FFCV \cite{ffcv}, which claimed a reduction of ImageNet model training up to 98 cents per model training. Furthermore, Deep Lake achieves similar ingestion performance to WebDataset \cite{webdataset}. Deep Lake significantly outperforms on larger images. Parquet is optimized for small cells and analytical workloads, while Deep Lake is optimized for large, dynamically shaped tensorial data. Compared to other data lake solutions, its minimal python package design enables Deep Lake to be easily integrated into large-scale distributed training or inference workloads.

\subsection{Future work}
The current implementation of Deep Lake has opportunities for further improvement. Firstly, the storage format does not support custom ordering for an even more efficient storage layout required for vector search or key-value indexing. Secondly, Deep Lake implements branch-based locks for concurrent access. Similar to Delta ACID transaction model \cite{delta}, Deep Lake can be extended to highly-performant parallel workloads. Thirdly, the current implementation of TQL only supports a subset of SQL operations (i.e., does not support operations such as \textit{join}). Further work will focus on making it SQL-complete, extending to more numeric operations, running federated queries in external data sources and benchmarking against SQL engines.

\section{Related Work}

Multiple projects have tried to improve upon or create new formats for storing unstructured datasets including TFRecord extending Protobuf \cite{Protobuf}, Petastorm \cite{Petastorm} extending Parquet \cite{parquet}, Feather \cite{feather} extending arrow  \cite{arrow}, Squirrel using MessagePack \cite{2022squirrelcore}, Beton in FFCV \cite{ffcv}. Designing a universal dataset format that solves all use cases is very challenging. Our approach was mostly inspired by CloudVolume \cite{cloudvolume}, a 4-D chunked NumPy storage for storing large volumetric biomedical data. There are other similar chunked NumPy array storage formats such as Zarr \cite{zarr}, TensorStore \cite{tensorstore}, TileDB \cite{papadopoulos2016tiledb}. Deep Lake introduced a typing system, dynamically shaped tensors, integration with fast deep learning streaming data loaders, queries on tensors and in-browser visualization support. An alternative approach to store large-scale datasets is to use HPC distributed file system such as Lustre \cite{schwan2003lustre}, extending with PyTorch cache \cite{kumar2020quiver} or performant storage layer such as AIStore \cite{aizman2019high}. Deep Lake datasets can be stored on top of POSIX or REST API-compatible distributed storage systems by leveraging their benefits. Other comparable approaches evolve in vector databases \cite{2021milvus, weaviate, 2021milvus} for storing embeddings, feature stores \cite{michelangelo, feast}  or data version control systems such as DVC \cite{ruslan_kuprieiev_2022_7039863}, or LakeFS \cite{lakefs}. In contrast, Deep Lake version control is in-built into the format without an external dependency, including Git. Tensor Query Language, similar to TQP \cite{he2022query} and Velox \cite{velox} approaches, runs n-dimensional numeric operations on tensor storage by truly leveraging the full capabilities of deep learning frameworks. Overall, Deep Lake takes parallels from data lakes such as Hudi, Iceberg, Delta \cite{delta, iceberg, hudi} and complements systems such as Databarick's Lakehouse \cite{lakehouse} for Deep Learning applications.

\section{Conclusion}
We presented Deep Lake, the lakehouse for deep learning. Deep Lake is designed to help deep learning workflows run as seamlessly as analytical workflows run on Modern Data Stack. Notably, Deep Lake is built to retain prominent features of data lakes, such as time travel, querying, and rapid data ingestion at scale. One important distinction from traditional data lakes is Deep Lake's ability to store unstructured data with all its metadata in deep learning-native columnar format, which enables rapid data streaming. This allows materializing data subsets on-the-fly, visualizing them in-browser, or ingesting them into deep learning frameworks without sacrificing GPU utilization. Finally, we show that Deep Lake achieves state-of-the-art performance for deep learning on large datasets via multiple benchmarks. 

\section{Acknowledgement}
The authors would like to thank Richard Socher, Travis Oliphant, Charu Rudrakshi, Artem Harutyunyan, Iason Ofeidis, Diego Kiedanski, Vishnu Nair, Fayaz Rahman, Dyllan McCreary, Benjamin Hindman, Eduard Grigoryan, Kristina Grigoryan, Ben Chislett, Joubin Houshyar, Andrii Liubimov, Assaf Pinhasi, Vishnu Nair, Eshan Arora, Shashank Agarwal, Pawel Janowski, Kristina Arezina, Gevorg Karapetyan, Vigen Sahakyan and the open-source community including contributors. The project was funded by Activeloop. We also thank the
CIDR reviewers for their feedback

\bibliographystyle{ACM-Reference-Format}
\bibliography{refs}

\end{document}
\endinput0